\documentclass[aps,prl,twocolumn,superscriptaddress,showpacs]{revtex4}

\usepackage{graphicx}

\begin{document}

\title{Hiding solutions in random satisfiability problems: A
statistical mechanics approach}

\author{W. Barthel}
\affiliation{Institute for Theoretical Physics, University of
G\"ottingen, Bunsenstr. 9, 37073 G\"ottingen, Germany}
\author{A.K. Hartmann}
\affiliation{Institute for Theoretical Physics, University of
G\"ottingen, Bunsenstr. 9, 37073 G\"ottingen, Germany}
\author{M. Leone}
\affiliation{SISSA, via Beirut 9, 34100 Trieste, Italy}
\affiliation{International Centre for Theoretical Physics, Strada
Costiera 11, P.O. Box 586, I-34100 Trieste, Italy}
\author{F. Ricci-Tersenghi}
\affiliation{International Centre for Theoretical Physics, Strada
Costiera 11, P.O. Box 586, I-34100 Trieste, Italy}
\affiliation{Dipartimento di Fisica, Universit\`a di Roma ``La
Sapienza'', Piazzale Aldo Moro 2, 00185 Roma, Italy}
\author{M. Weigt}
\affiliation{Institute for Theoretical Physics, University of
G\"ottingen, Bunsenstr. 9, 37073 G\"ottingen, Germany}
\author{R. Zecchina}
\affiliation{International Centre for Theoretical Physics, Strada
Costiera 11, P.O. Box 586, I-34100 Trieste, Italy}

\date{\today}

\begin{abstract}
A major problem in evaluating stochastic local search algorithms for
NP-complete problems is the need for a systematic generation of hard
test instances having previously known properties of the optimal
solutions. On the basis of statistical mechanics results, we propose
random generators of hard and satisfiable instances for the
3-satisfiability problem (3-SAT). The design of the hardest problem
instances is based on the existence of a first order ferromagnetic
phase transition and the glassy nature of excited states. The
analytical predictions are corroborated by numerical results obtained
from complete as well as stochastic local algorithms.
\end{abstract}


\maketitle

In natural sciences and in artificial systems, there exist many
problems whose solution requires computational resources growing
exponentially with the number of variables $N$ needed for their
encoding. Concrete examples are optimization and cryptographic
problems in computer science, glassy systems and random structures in
physics and chemistry, random graphs in mathematics, and scheduling
problems in real-world applications.

Having fast and powerful algorithms for the resolution of these
problems is of primary relevance for their theoretical study as well
as for applications.  The evaluation of such algorithms is based on
the availability of hard benchmarks, having the following properties:
They provide problem instances, with a given known solution, in a fast
way (e.g.\ linear in $N$), but the resolution of an instance takes a
time exponential in $N$ for any known algorithm. So the best
algorithms can be easily selected. In this letter we propose a new
generator of hard and solvable test instances, having all the
properties listed above.  It is based on a NP-complete problem
\cite{GaJo} -- namely 3-satisfiability (3-SAT).

The main idea for the construction of such hard and solvable problems
is very simple: to hide a known solution within a multitude of
coexisting random meta-stable configurations which constitute
dynamical barriers.  In the physical approach based on a mapping from
3-SAT to a spin glass model \cite{MoZe}, such random configurations
correspond to glassy states \cite{FLRZ}. It is to be noted, however,
that many previous attempts to implement this idea were unsuccessful,
because the random structure was usually easy to remove, or knowledge
that a solution has been forced can be exploited to find it. In the
instances we propose, instead, the presence of a known solution does
not alter the structure of the glassy state, which confuses the solver
and makes the problem hard.

As an important application of these ideas to cryptography
\cite{crypto}, random one-way functions are provided: A given message,
e.g.\ a password, can be coded in a 3-SAT formula and thus verified
efficiently, but decoding it is extremely time-consuming.

We use the framework of the typical-case computational complexity
\cite{AI,TCS}. There, the study of random 3-SAT problems has played a
major role. A random 3-SAT formula $F$ consists of $M$ logical clauses
$\{C_\mu\}_{\mu=1,...,M}$ over a set of $N$ Boolean variables $\{x_i =
0,1\}_{i=1,...,N}$, with 0=FALSE and 1=TRUE.  Every clause consists of
three randomly chosen Boolean variables which are connected by logical
OR operations ($\vee$) and appear negated with probability 1/2, e.g.\
$C_\mu=( x_i \vee \overline{x}_j \vee x_k)$. In $F$ the clauses are
connected by logical AND operations ($\wedge$), $F=\bigwedge_{\mu=1}^M
C_\mu\ $, so that all clauses have to be satisfied simultaneously in
order to satisfy the formula.

A satisfying logical assignment of the $x_i$ is also called a solution
of $F$.  The random 3-SAT model was found to undergo a SAT/UNSAT phase
transition \cite{MiLeSe} at a critical ratio $\alpha_c = M/N \simeq
4.25$ ($N\gg1$): Below $\alpha_c$, almost all formulae are
satisfiable, while beyond almost all formulae do not show any
solution. At this threshold, a strong exponential peak in the typical
(median) cost for finding solutions by the best known algorithms
appears. Problem instances generated close to it form a natural test
bed for the optimization of heuristic search algorithms. However,
satisfiable and unsatisfiable instances coexist in this region. Many
algorithms of practical interest \cite{satlib} are based on {\em
incomplete} stochastic local search procedures, as e.g.\ simulated
annealing \cite{SIM_ANN} and the walk-SAT algorithm \cite{Se}. These
algorithms stop once they have found a solution, but they have no way
to disentangle, in polynomial time in $N$, if a formula is
unsatisfiable or just hard to solve. It is thus very important to
generate benchmarks which are satisfiable and for which the
algorithmic proof of this satisfiability takes an exponential time in
N.

In this Letter, we propose simple and fast generators of such
benchmark problems. The main ideas are inspired by physical
requirements, and exploit the presumed hardness of random 3-SAT
itself. One obvious possibility \cite{satlib} is to filter the
problems at the phase boundary by complete algorithms, and to keep
only the satisfiable ones. This method is limited by the small values
of $N$ and $M$ which can be handled by the filtering algorithms, thus
making the generation itself exponentially long.  In addition, the
hardest instances are the unsatisfiable ones.  Other approaches use
mappings from various hard problems to 3-SAT, including e.g.\
factorization \cite{HoWa}, graph coloring \cite{HoSt}, and Latin
square completion \cite{AcKaGoSe}.

We choose an arbitrary assignment of our logical variables and accept,
with some prescribed probability, only clauses which are satisfied by
this assignment. Without loss of generality, we restrict ourselves to
generating formulae which are satisfied by $x^{(0)}_i=1,\ \forall
i=1,..,N$ \cite{note1}. So, only clauses containing three negated
variables are excluded; all other clauses are satisfied by $\vec
x^{(0)}$.  The generation of random 3-SAT formulae is done as follows:
For each of the $M=\alpha N$ clauses, we draw randomly and
independently three indices $i,j,k \in \{1,...,N\}$. Then, we choose
one of the seven allowed clauses with the following probabilities:
clause $(x_i \vee x_j \vee x_k)$ -- type ``0'' -- with probability
$p_0$; each of the clauses $(\overline{x}_i \vee x_j \vee x_k)$, $(x_i
\vee \overline{x}_j \vee x_k)$ and $(x_i \vee x_j \vee
\overline{x}_k)$ -- type ``1'' -- with probability $p_1$; and finally
each of $(\overline{x}_i \vee \overline{x}_j \vee x_k)$,
$(\overline{x}_i \vee x_j \vee \overline{x}_k)$ and $(x_i \vee
\overline{x}_j \vee \overline{x}_k)$ -- type ``2'' -- with probability
$p_2$, where $p_0+3 p_1+3 p_2=1$. As we will show in the following,
typically hard instances can be generated if the parameters are chosen
as follows:
\begin{eqnarray}
\alpha > 4.25\quad , & \quad & 0.077 < p_0 < 0.25\; , \nonumber \\
p_1 = (1-4p_0)/6\; , & \quad & p_2 = (1+2p_0)/6 \;\;\: . \label{eq1}
\end{eqnarray}
To understand this model, and to find values for $p_0,\ p_1$ and $p_2$
such that the instances are as hard as possible, we have followed a
statistical mechanics approach corroborated by numerical simulations
based on both complete and randomized algorithms.  The analysis is
based on the standard representation of 3-SAT as a diluted spin-glass
model \cite{MoZe}: The Boolean variables $x_i=0,1$ are mapped to Ising
spins $S_i=(-1)^{x_i}$, and the Hamiltonian counts the number of
unsatisfied clauses,
\begin{equation}
\mathcal{H} = \frac{\alpha}{8} N - \sum_{i=1}^N H_i S_i - \sum_{i<j}
T_{ij} S_i S_j - \sum_{i<j<k} J_{ijk} S_i S_j S_k
\label{eq:ham}
\end{equation}
with $H_i=\frac{1}{8}\sum_\mu c_{\mu,i}$, $T_{ij} = -\frac{1}{8}
\sum_\mu c_{\mu,i} c_{\mu,j}$, and $J_{ijk}= \frac{1}{8} \sum_\mu
c_{\mu,i} c_{\mu,j} c_{\mu,k}$, where $c_{\mu,i}$ equals $+1$ if $x_i$
appears directly in $C_\mu$, $-1$ if it appears negated, and $0$
otherwise. The interactions in (\ref{eq:ham}) fluctuate from sample to
sample, with disorder-averages $\overline{H_i} = \frac{3\alpha}{8}
(p_0+p_1-p_2)$, $\overline{T_{ij}} = \frac{3\alpha}{4N}
(-p_0+p_1+p_2)$, and $\overline{J_{ijk}} = \frac{3\alpha}{4N^2}
(p_0-3p_1+3p_2)$.

We are interested in the ground states of this Hamiltonian. For a
satisfiable formula we know that the corresponding ground state energy
vanishes. In order to analytically characterize the ground states
properties, we first calculate the free energy at formal temperature
$T$, using the functional replica trick in the replica-symmetric
framework \cite{MoZe}. Then we send $T \to 0$, and we study the
zero-temperature phase diagram of the model using $\alpha$ and
$p_{0,1,2}$ as control parameters.

The replica-symmetric order parameter determining the different phases
of the system is the distribution of local magnetizations $P(m)=1/N\
\sum_i \delta(m-m_i)$, where $m_i=\langle S_i \rangle_{T=0}$ is the
average value of $S_i$ over all ground states.  There are mainly two
different cases:\\
(\textit{i}) $P(m)$ has a non-zero average and/or is broad, but all
$|m_i|$ are less than 1. It can be determined using a simple
population dynamics algorithm \cite{MePa} or variationally
\cite{BiMoWe}. Both results coincide.\\
(\textit{ii}) $P(m)$ can be calculated exactly and turns out to have a
finite weight in $m = 1$, i.e.\ an extensive number of variables is
fixed to $x_i=1$ in all satisfying assignment (the so called {\it
backbone} \cite{nature}).

Going back to the class of generators proposed above, one could
naively use $p_0=p_1=p_2=1/7$ (\textit{model~1/7}), choosing any of
the allowed clauses with the same probability. This generator,
including some extensions, \cite{AsIwMi,MaHo,MoUe} is known to be
effectively solvable by local search procedures \cite{AcKaGoSe}. In
our walk-SAT implementation, the maximal resolution-time \cite{note2}
grows like $t\propto N^{1.58}$, and large systems of sizes up to $N
\simeq 10^4$ can be easily handled.

The statistical mechanics approach clarifies this result: The proposed
generator behaves like a paramagnet in an exterior random field, and
no ferromagnetic phase transition appears. Local search algorithms may
exploit the average local field $\overline{H_i} = 3\alpha/56$ pointing
into the direction of the forced solution $\vec x^{(0)}$, and rapidly
find a solution.

To avoid this, we can fix the average local field to zero by choosing
$p_0+p_1-p_2=0$. The probabilities are thus restricted by $0 \le p_0
\le 1/4$, $p_1=(1-4p_0)/6$, and $p_2=(1+2p_0)/6$.

\begin{figure}
\includegraphics[width=\columnwidth]{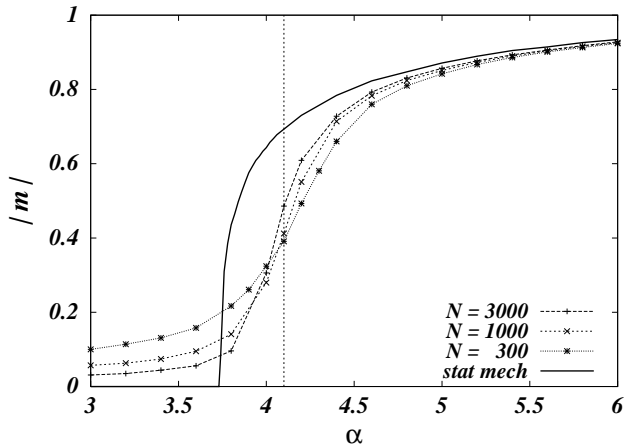}
\caption{Magnetization of the first walk-SAT solution in model 1/6.
Due to the (average) spin-flip symmetry, we plot the average of $|m|$.
For large $N$, the magnetization stays zero up to $\alpha \simeq
4.1$. The full line shows the thermodynamic average, which stays well
above the asymptotic walk-SAT result.}
\label{walksat1/6mag}
\end{figure}

Let us start the discussion of these possibilities with the case
$p_0=0$, $p_1=p_2=1/6$ (\textit{model 1/6}). In this (and only this)
case, there is a second guaranteed solution: $x_i=0,\ \forall i$. The
average $\overline{J_{ijk}}$ vanishes, too. The model is paramagnetic
at low $\alpha$, and undergoes a second order ferromagnetic transition
at $\alpha \simeq 3.74$ (see full line in
Fig.~\ref{walksat1/6mag}). But also in the ferromagnetic phase the
backbone is still zero as long as $\alpha \alt 4.91$: At this point it
appears continuously from strongly magnetized spins.

\begin{figure}
\includegraphics[width=\columnwidth]{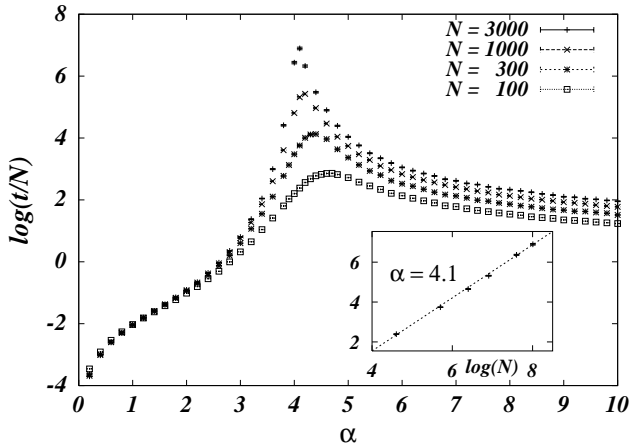}
\caption{Typical walk-SAT complexity for model 1/6.  We show the
average value of $\log(t/N)$.  We find a clear data collapse for small
$\alpha$ in the linear regime, $t \propto N$.  The complexity peak at
$\alpha \simeq 4.1$ grows polynomially as shown in the inset. The
slope of the straight line in the inset is 1.3.}
\label{walksat1/6}
\end{figure}

In walk-SAT experiments, we find that the generated instances are
still solvable in polynomial time, with peak resolution-times growing
as $N^{2.3}$, see Fig.~\ref{walksat1/6}.  However, the complexity peak
is not at the phase transition, but quite close to the critical point
of random 3-SAT. This is due to the fact that walk-SAT does not sample
solutions according to the thermodynamic equilibrium distribution:
Most probably it hits solutions with small magnetization, i.e.\ closer
to the starting point (see Fig.~\ref{walksat1/6mag}).  For $N\to
\infty$, this magnetization stays zero even after the ferromagnetic
transition. Indeed, if we restrict the statistical mechanics analysis
to zero magnetization, we find an exponential number of solutions also
beyond $\alpha=3.74$. More interestingly, this number coincides with
the one of random 3-SAT, which jumps to 0 at $\alpha\simeq 4.25$
\cite{MoZe}. So, approaching this point, walk-SAT is no longer able to
find unmagnetized solutions for model 1/6, and it has to go to
magnetized assignments, giving rise to the resolution-time peak.

Once we use $p_0>0$, the situation changes: The ferromagnetic
transition becomes first order, as can be seen best by the existence
of metastable solutions for $P(m)$. The transition point moves towards
the random 3-SAT threshold $\alpha_c$, and the computational
complexity increases with $p_0$. Still, for $p_0 \alt 0.077$, the
ferromagnetic phase arises without backbone and solutions can be
easily found.

\begin{figure}
\includegraphics[width=\columnwidth]{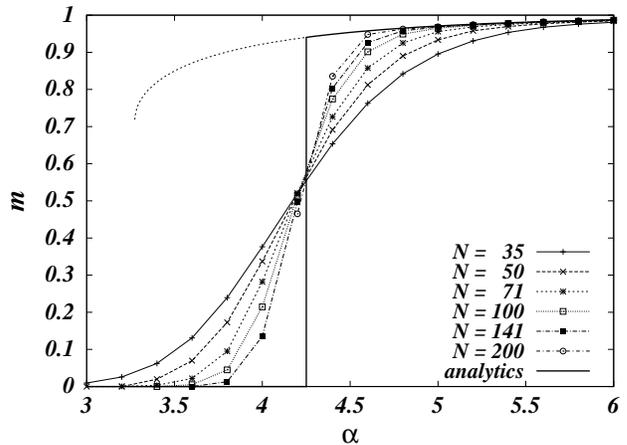}
\caption{Average magnetization of solutions of model 1/4, obtained
with a complete algorithm. There, the magnetization equals the
backbone size. The finite size curves cross at $\alpha \simeq 4.25$,
and tend to the analytical prediction.  The dotted continuation of the
analytical line gives the globally unstable ferromagnetic solution,
starting at the spinodal point.}
\label{dp1/4mag}
\end{figure}

In the region $0.077 \alt p_0 < 1/4$, the first order transition is
more pronounced. The system jumps at $\alpha\simeq 4.25$ from a
paramagnetic phase to a ferromagnetic one, with a discontinuous
appearance of a backbone: For $p_0\simeq 0.077$, the backbone size at
the threshold is about $0.72 N$, and goes up to $0.94 N$ for $p_0=1/4$
(see Fig.~\ref{dp1/4mag}).  We conjecture the ferromagnetic critical
point in these models to coincide with the SAT/UNSAT threshold in
random 3-SAT, since the topological structures giving rise to
ferromagnetism in the formers induce frustration and thus
unsatisfiability in the latter.

The case $p_0\!=\!1/4$, and so $p_1\!=\!0,\ p_2\!=\!1/4$
(\textit{model~1/4}), is very peculiar because it can always be solved
in polynomial time using a global algorithm. Indeed, one can
unambiguously add three clauses to every existing one, namely the
other clauses allowed in model 1/4, without loosing the satisfiability
of the enlarged formula \cite{note3}.  The completed formula becomes a
sample of random satisfiable 3-XOR-SAT (also known as hyper-SAT
\cite{RiWeZe}), which can be mapped to a system of linear equations
modulo 2, and solved in time of $\mathcal{O}(N^3)$ \cite{note4}.

This algorithm immediately breaks down if we choose $p_0 \neq 1/4$.
Indeed, whenever one tries to map the general formula into a completed
one, the presence of all three types of clauses forces it into a {\em
frustrated} 3-XOR-SAT formula, which undergoes a SAT/UNSAT transition
at $\alpha = 0.918$ \cite{RiWeZe}, well below the region of our
interest. So the mapping is of no use for $p_0 \neq 1/4$.  In this
case, any 3-SAT instance with solution $\vec x^{(0)}$ (and thus any
solvable one \cite{note1}) can be generated with non-zero probability.
The worst-case is thus included in the presented generator, and there
cannot be any polynomial solver if P$\neq$NP.

\begin{figure}
\includegraphics[width=1.0\columnwidth]{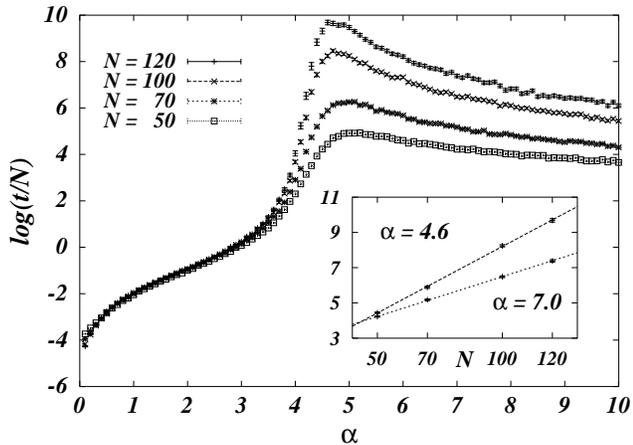}
\caption{Typical walk-SAT complexity for model 1/4. The complexity
peak is much more pronounced than in Fig.~\ref{walksat1/6}, cf.\ e.g.\
the reachable system sizes. The inset shows the exponential
resolution-time scaling near the peak ($\alpha=4.6$) and deep inside
the ferromagnetic phase ($\alpha=7.0$). The slopes of the straight
lines are 0.075 and 0.04 respectively.}
\label{walksat1/4}
\end{figure}

In the following table we summarize the main results for the
investigated combinations of $p_0,\ p_1$ and $p_2$.  Where only $p_0$
is reported, $p_{1,2}$ are given by Eqs.(\ref{eq1}).  We show the
location $\alpha_c$ and order of the ferromagnetic phase transition,
together with the point $\alpha_{ws}$ and the system-size-scaling
(P/EXP) of the maximal walk-SAT complexity.  For comparison, we have
added the corresponding data for random 3-SAT.

\begin{center}
\begin{tabular}{||l|c|l||}\hline\hline
Model           & $\alpha_c$ (order, type)      & $\alpha_{ws}$ \\ 
\hline \hline
$p_{0,1,2}= 1/7$        & NO                    & 5.10 P \\ \hline
$p_0 = 0$               & 3.74 (2nd, ferro)     & 4.10 P \\ \hline
$p_0 \in [0.077, 1/4)$  & 4.25 (1st, ferro)     & 4.25 EXP \\ \hline
$p_0 = 1/4$             & 4.25 (1st, ferro)     & 4.25 P \\ 
\hline \hline
Random 3-SAT            & 4.25 (SAT/UNSAT)      & 4.25 EXP \\ 
\hline\hline 
\end{tabular}
\end{center}

Please note, that the polynomial time-complexity of model 1/4 is
accidental and due to the existence of a global algorithm, whereas the
walk-SAT peak grows exponentially with $N$. To corroborate this
picture, we also performed simulated annealing experiments. We easily
find solutions in model 1/6, but get stuck in the vicinity of model
1/4.

As a conclusion, we conjecture the hardest instances to be generated
with $p_0$ values close to 1/4. The computational times for their
solution are similar to those in Fig.~\ref{walksat1/4}, which have
been obtained for $p_0=1/4$ without exploiting the global algorithm.
Resolution-times are clearly exponential in all the ferromagnetic
phase ($\alpha > 4.25$).  Moreover we checked that resolution-times in
the paramagnetic phase ($\alpha < 4.25$) coincide, up to finite-size
effects, with those of random 3-SAT.

The physical interpretation of the hardness in this class of models is
based on the presence of glassy metastable states of zero
magnetization \cite{FLRZ} for $\alpha>4.25$.  These states are
dynamically favored and trap the system for very long times during a
stochastic local search.  We believe that the statistical mechanics
approach can have a general valence in the formulation of hard and
solvable problems, allowing for a systematic way of producing random
one-way functions, and can help in the study of the dynamics of
randomized search algorithms.

\begin{acknowledgments}
MW thanks the ICTP in Trieste and RZ thanks the LPTMS in Orsay for
hospitality.
\end{acknowledgments}


\begin{thebibliography}{99}
\bibitem{note1} The satisfying assignment $\vec x^{(0)}$ can be
changed to any other one $\vec x^{(1)}$ by a local gauge
transformation of the generated 3-SAT formula: One has to exchange
$x_i$ and $\overline x_i$ for all $i$ with $x_i^{(1)}=0$.
\bibitem{note2} The resolution-time $t$ is measured by the number of
single-variable flips until a satisfying solution is found.
\bibitem{note3} In the completed model, every clause is replaced by
groups of four in the form $(x_i \vee x_j \vee x_k) \wedge (x_i \vee
\bar x_j \vee \bar x_k) \wedge (\bar x_i \vee x_j \vee \bar x_k)
\wedge (\bar x_i \vee \bar x_j \vee x_k) = (x_i \oplus x_j \oplus
x_k)$ with $\oplus$ being the logical XOR operator.  One is left with
a ferromagnetic pure 3-spin model, see \cite{RiWeZe}.
\bibitem{note4} Note, however, that in the worst case finding a
solution of zero magnetization in the model 1/4 is NP-complete (the
proof will be given in a forthcoming paper).
\end{thebibliography}
\end{document}